\documentclass{aa}
\usepackage{graphicx}
\usepackage{txfonts}
\usepackage{natbib}
\usepackage{url}

\newcommand{\arsec}{'\!\hskip0.4pt'\hspace{-0.2mm}}

\newcommand{\dotarsec}{.\hspace{-0.9mm}'\!\hskip0.4pt'\hspace{-0.2mm}}

\begin{document}
   \title{The continuum intensity as a function of magnetic field}
   \subtitle{II. Local magnetic flux and convective flows }

    \author{P. Kobel \inst{1,2} \and S. K. Solanki \inst{1,3} \and J. M. Borrero \inst{4} }

   \offprints{P. Kobel}

   \institute{Max-Planck Institut f\"{u}r Sonnensystemforschung, Max-Planck-Stra\ss e 2, 37191 Katlenburg-Lindau, Germany\\
              \email{philippe.kobel@a3.epfl.ch}
              \and EPFL, Laboratoire des Machines Hydrauliques, 1007 Lausanne, Switzerland \\
              \and School of Space Research, Kyung Hee University, Yongin, Gyeonggi, 446-701, Korea \\
              \and Kiepenheuer-Institut f\"{u}r Sonnenphysik, Sch\"{o}neckstr. 6, 79104 Freiburg, Germany }

   \date{...}


  \abstract
      {To deepen our understanding of the role of small-scale magnetic fields in active regions (ARs) and in the
  quiet Sun (QS) on the solar irradiance, it is fundamental to investigate the physical processes underlying
  their continuum brightness. Previous results showed that magnetic elements in the QS reach larger continuum
  intensities than in ARs at disk center, but left this difference unexplained.}
      {We use Hinode/SP disk center data to study the influence of the local
      amount of magnetic flux on the vigour of the convective flows and the continuum intensity contrasts.}
      {The apparent (i.e. averaged over a pixel) longitudinal field strength and line-of-sight (LOS) plasma
  velocity were retrieved by means of Milne-Eddington inversions (VFISV code). We analyzed a series of boxes
  taken over AR plages and the QS, to determine how the continuum intensity contrast of magnetic elements, the
  amplitude of the vertical flows and the box-averaged contrast were affected by the mean longitudinal field
  strength in the box (which scales with the total unsigned flux in the box).}
      {Both the continuum brightness of the magnetic elements and the dispersion of the LOS velocities
      anti-correlate with the mean longitudinal field strength. This can be attributed to the ``magnetic
      patches'' (here defined as areas where the longitudinal field strength is above 100 G) carrying most of the
      flux in the boxes. There the velocity amplitude and the spatial scale of convection are reduced. Due to
      this hampered convective transport, these patches appear darker than their surroundings. Consequently, the
      average brightness of a box decreases as the the patches occupy a larger fraction of it and the amount of
      embedded flux thereby increases.}
      {Our results suggest that as the magnetic flux increases locally (e.g. from weak network to strong plage),
      the heating of the magnetic elements is reduced by the intermediate of a more suppressed convective energy
      transport within the larger and stronger magnetic patches. This, together with the known presence of larger
      magnetic features, could explain the previously found lower contrasts of the brightest magnetic elements in
      ARs compared to the QS. The inhibition of convection also affects the average continuum brightness of a
      photospheric region, so that at disk center, an area of photosphere in strong network or plage
      appears darker than a purely quiet one. This is qualitatively consistent with the predictions of 3D MHD
      simulations.}


   \titlerunning{The continuum intensity as a function of magnetic field II.}
   \authorrunning{P. Kobel et al.}

   \maketitle
%

\section{Introduction}
\label{sec_intro}

Variations of the Total Solar Irradiance (TSI) on timescales from days to the solar cycle can be
reasonably well reproduced based on the evolving distribution of solar surface magnetic fields \citep[which is
the basis of e.g. the SATIRE reconstructions,][]{Fligge00, Solanki02, Krivova03, Wenzler06}. Although the TSI
reconstructions proved very successful \citep[with a correlation of up to 0.98 with the measured
irradiance,][]{Ball11}, they are based on an empirical relationship between the emergent intensity of a pixel and
the measured magnetic field (via MDI magnetograms) whose simple linearity lacks physical basis; its justification
relies on its ability to reproduce the disk-average irradiance and the convenience of using a single model
atmosphere for all faculae \citep[for more details see][]{Fligge00}. In particular, the spectral radiance of a
given pixel is assumed to be \emph{uniquely} determined by the magnetic signal at that pixel, thus neglecting its
magnetic environment and location on the Sun.

The surface magnetic field outside Sunspots is mainly distributed in active region (AR) plages and in the quiet
Sun (QS) network outlining supergranular cells. There the magnetic flux is often concentrated into features
traditionally described in terms of flux tubes, with field strengths of the order of kG \citep[e.g.][]{Frazier72,
Stenflo73, Rabin92a, Ruedi92} and a spectrum of sizes \citep[see][for a review]{Solanki_rev06}. At the lower end
of this size spectrum are found the so-called ``magnetic elements'' \citep[see][]{Schuessler_rev92,
Solanki_rev93}, which have been spatially resolved recently by the balloon-borne SUNRISE ImaX observations at an
angular resolution of $\sim 0\dotarsec14$ \citep[][]{Lagg10}. As a result of being hotter than their surroundings
at equal optical depth \citep{Schuessler_rev92}, they often appear brighter than the average quiet photosphere,
i.e. they have a positive ``contrast''. This is particularly pronounced when observing in the core of spectral
lines and in molecular bands \citep{Chapman68, Sheeley69, Muller84}, but also holds at continuum wavelengths or
broader wavelength bands, even at disk center, if the magnetic elements are sufficiently resolved \citep[see][the
two last citations refering to radiation-MHD simulations]{Muller83b, Foukal84, Lawrence88, Riethmueller10,
Schuessler88, Voegler03}. Note that the contrast of magnetic features further increases from the disk center to
the limb in a way that is still a matter of debate \citep{Steiner07} and shall be treated in a forthcoming paper.
Therefore,
taking into account the larger area coverage of magnetic elements compared to sunspots, their radiance
overcompensates the solar brightness deficit due to sunspots at activity maximum \citep{Froehlich00}. It is known
that ARs typically harbour more magnetic flux per unit area than the network and thus a higher number density of
magnetic elements. The larger available flux in ARs also leads to the formation of larger and darker magnetic
features such as pores and micropores, the latter being small (sub-arcsec) magnetic features that are darker than
the QS continuum \citep[][]{Beckers68, Tarbell77}. The inner part of supergranulation cells, the so-called
``internetwork'' (IN), also contains small magnetic elements but in lower number density \citep[][]{Muller83a,
Lites02, Dominguez03, deWijn05, deWijn08}. Finally, the solar photosphere contains weaker equipartition fields
everywhere \citep[detected mostly in the IN by, e.g.][]{Lin95, Solanki96, Lites02, Khomenko03}, but the latter
have been estimated to bring negligible contributions to the TSI variations \citep{Schnerr10}.

To determine how these different components of solar surface magnetism contribute to the solar irradiance, one
should investigate the intensity-magnetic field relation within ARs, the QS network and even the IN. In the
1990's, \citet{Title89, Topka92, Lawrence93} and \citet{Topka97} performed a series of ground-based studies of
continuum intensity vs. magnetogram signal in ARs and in the QS at disk center, and found that the QS continuum
intensity contrast reaches larger values than in ARs at equal magnetogram signal. This early result was confirmed
by our recent study using Hinode/SP data at a constant and higher spatial resolution than the earlier studies
\citep[][hereafter Paper I]{Kobel11a}. Note that in our study, the contrast of magnetic elements could be
interpreted as a measure of their intrinsic brightness by virtue of the comparable contrast references used in
ARs and in the QS. We also found that the bright magnetic elements in ARs and in the QS share similar filling
factors (along with similar kG field strengths and inclinations close to vertical). Assuming that, at Hinode's
resolution and at disk center, the filling factor reflects the size of the unresolved magnetic elements (at the
height of line formation), this questions the conventional interpretation that the brightness of magnetic
features is \emph{primarily} dictated by their size \citep[][see also the Introduction of Paper I]{Spruit81}.

As discussed in Paper I, the other factor that could possibly influence the brightness of magnetic elements
(besides their size) is the surrounding convective energy transport, because it determines the energy available
to be radiated into the flux tubes through their ``hot walls'' \citep[][]{Spruit76}. Since AR plages typically
have a larger mean field strength than the QS network, the Lorentz force should inhibit the convective flows more
strongly in ARs. The granulation indeed appears ``abnormal'' in plages \citep[e.g.][]{Dunn73}, such that its
spatial scale is reduced compared to the QS, as are both the vertical and horizontal components of the
flow field \citep{Schmidt88, Title89, Title92, Keller91}. 3D MHD simulations also disclose a similar behaviour as
the mean field increases \citep{Voegler05b}. Furthermore, these carefully set up simulations (with constant
inflowing entropy density at the bottom boundary) demonstrated that the convective energy transport was
increasingly inhibited to the point of altering the \emph{vertical} radiative energy output from the boxes,
thus providing an explanation for magnetic elements appearing darker in more active environments at disk
center.

The work presented herein is an attempt to explain the effect of the local amount of magnetic flux on the
continuum brightness of magnetic elements and on the local average continuum brightness of the solar surface
at disk center. Inspired by MHD simulations, we carried out a ``local box analysis'' that considers a
series of small square fields of view taken from different regions of QS and ARs (see Sect. \ref{sec_boxes})
located at disk center, observed with the spectropolarimeter instrument onboard Hinode. In Sect.
\ref{sec_results}, we examine the relation between the contrast and the longitudinal field strength inside these
boxes. In particular, we assess the influence of the mean longitudinal field strength in the box on the intensity
contrast of the magnetic elements, on the vigour of the vertical convective flows, and on the contrast averaged
over the box. One reason to restrict our study to disk center is that in this way we can interpret the
Doppler velocities retrieved by Milne-Eddington inversions from the data (see Sect.\ref{sec_scans_inversions}) in
terms of vertical flows. These results are then discussed in Sect. \ref{sec_disc}, with special emphasis on the
role of the magnetically-suppressed convection on the contrasts.


\section{Dataset}

\subsection{Hinode/SP scans and inversions}
\label{sec_scans_inversions}

We used an ensemble of 6 spectropolarimetric scans over ARs and 4 scans over the QS performed
very close to disk center (see Table \ref{table_scans}) by the Hinode/SP instrument in its ``normal mode'',
corresponding to an integration time of 4.8 s and rms polarimetric noise of $\sim 10^{-3}$ in units of the
continuum intensity $I_c$ \citep{HinodeSOT, HinodeSOT2}. Maps of $I_c$ and $\mu$ values (calculated in the red
continuum of the 630.2 nm line) were calculated by the \verb"sp_prep" procedure. Note that we used the same scans
as in Paper I.

The observed Stokes spectra at each pixel of the scans were inverted with the VFISV (Very Fast Inversion of
the Stokes Vector) Milne-Eddington code of \citet{Borrero10} (we refer to that article for all details of the
code). Unlike in Paper I, no filling factor treatment was considered here, as we were only interested in
retrieving quantities \emph{averaged over a pixel}.\footnote{This means that these quantities were inferred by
taking into account all the light detected at the pixel, as if it were entirely coming from a magnetic
atmosphere. These photons actually originate from the solar area sampled by the pixel (which can be only
partially covered by magnetic fields) and from its local neighbourhood \citep{Orozco07} due to diffraction by the
relatively narrow point spread function of the Hinode/SOT \citep{Danilovic08}.} In particular, we were concerned
by the following inversion parameters: the line-of-sight (LOS) components of the plasma velocity, $v_{\rm los}$,
and the apparent strength of magnetic field $B_{\rm app,los} = B |$cos$\gamma|$, where $B$ is the field strength
and $\gamma$ the inclination with respect to the LOS. Since $B$ is averaged over the pixel (and thus not
``intrinsic''), this definition of $B_{\rm app,los}$ is physically equivalent to the one used in Paper I, which
is the reason for keeping the same notation and terminology (``apparent'') for it.

\begin{table}
\label{table_scans}
\caption{List of the SP scans used in this work. $t_{\rm start}$ denotes the starting time of the
scans, and ``target'' indicates whether the scan was performed over an active region (AR), or the quiet Sun (QS).}
\vskip3mm \centering
\begin{tabular}{c c c c c}     
\hline\hline
date (dd-mm-yy) & $t_{\rm start}$ (UT) & target & NOAA & No. boxes \\
\hline
11-12-06 & 13:10:09 & AR & 10930 & 3\\
05-01-07 & 11:20:09 & AR & 10933 & 5\\
01-02-07 & 12:14:05 & AR & 10940 & 5\\
28-02-07 & 11:54:34 & AR & 10944 & 2\\
01-05-07 & 21:00:06 & AR & 10953 & 7\\
11-05-07 & 12:35:53 & AR & 10955 & 3\\
\hline
10-03-07 & 11:37:36 & QS & -- & 5\tablefootmark{s} / 2\tablefootmark{w}\\
23-04-07 & 11:14:06 & QS & -- & 5\tablefootmark{s} / 1\tablefootmark{w}\\
24-04-07 & 01:21:04 & QS & -- & 2\tablefootmark{s} / 1\tablefootmark{w}\\
27-04-07 & 08:50:03 & QS & -- & 4\tablefootmark{s} / 1\tablefootmark{w}\\
\hline
\end{tabular}
\tablefoot{The number of boxes (``No. boxes'') extracted from each scan depended on the effective area of the
field of view without sunspots and with $\mu > 0.99$.\\
\tablefoottext{s}{Number of strong network boxes.}
\tablefoottext{w}{Number of weak network boxes.}}
\end{table}

\subsection{Local boxes, contrast definition and velocity calibration}
\label{sec_boxes}

\begin{figure*}
\centering
\includegraphics[width=\textwidth]{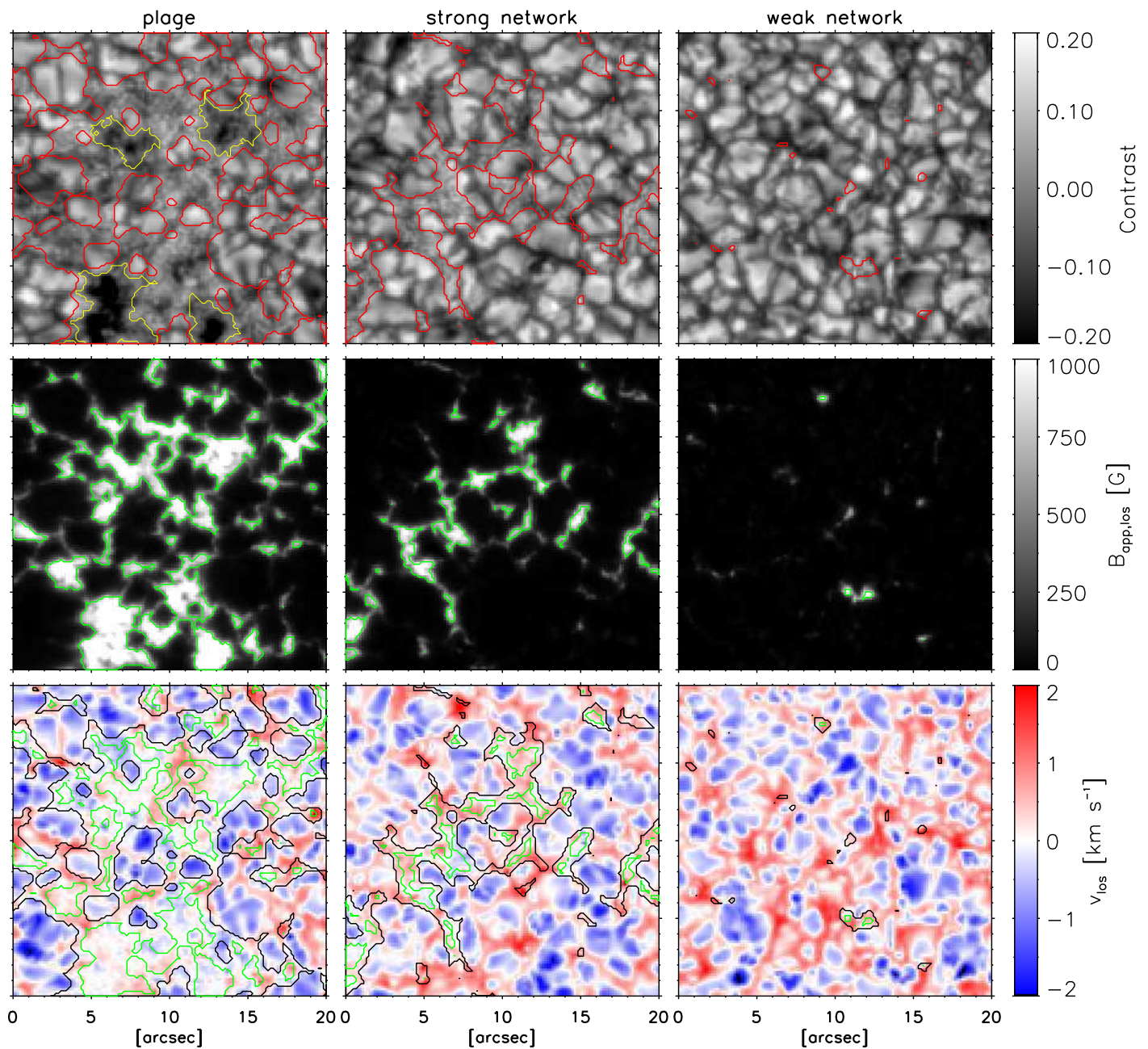}
\caption{Continuum contrast (top), apparent longitudinal field strength $B_{\rm app,los}$ (middle) and LOS
velocity $v_{\rm los}$ (bottom) for three examples of boxes in active region plage (left), strong network
(middle) and weak network (right), respectively. All the boxes are located at disk center ($\mu > 0.99$). The
weak network box has been cropped to a subfield of $20\arsec \times 20\arsec$ for better visibility, although its
original size is $70\arsec \times 70\arsec$. Yellow contours in top left: pores (see main text) excluded from the
scatterplot of contrast vs. $B_{\rm app,los}$ (Fig. \ref{fig_C_vs_B_boxes}) and from the averaged contrasts (Fig.
\ref{fig_avgC_vs_avgB}). Red contours (top) and black contours (bottom): delimit the areas where $B_{\rm app,los}
> 100$ G (magnetic patches) from the weakly magnetized areas whose mean intensity is used as reference for the
contrast (see Eq. \ref{eq_contrast}). Green contours (middle and bottom): delimit the location of the pixels
where $B_{\rm app,los} > 500$ G, so that the region between black and green contours has 100 G $< B_{\rm
app,los} < 500$ G.}
\label{fig_boxes}
\end{figure*}

From the different AR and QS scans we extracted a series of square ``boxes'' containing different
amounts of magnetic flux, all at $\mu > 0.99$.
In the network scans we found it useful to visually distinguish between ``strong network'', where the large
magnetic patches are reminiscent of plages, and ``weak network'' containing rather isolated magnetic elements.
Note that the strong network is similar to what has been called ``enhanced network''. The size of the boxes was a
trade-off between having enough statistics to investigate the contrast of their magnetic elements, while being
small enough to avoid mixing the three categories, namely plage, strong network and weak network. For the AR
plage and the QS strong network we chose a box size of $20\arsec \times 20\arsec$. Due to the low concentration of
flux and consequently poorer statistics in the weak network , we treated it in larger boxes of $70\arsec \times
70 \arsec$. Note that most of the area of the weak network boxes is covered by internetwork fields. These
contribute significantly to the total flux in these boxes. Also, it is now established that at least some of the
magnetic features in the IN have kG field strengths and correspond to magnetic elements \citep{Lagg10,
Sanchez03}. We selected a total of 25 plage boxes, 16 strong network boxes and 4 weak network/internetwork boxes.
One example of each category is provided in Fig. \ref{fig_boxes} (where the weak network/internetwork boxes was
cropped to an area of $20\arsec$ for comparison).

We quantified the total unsigned magnetic flux in each box, $\Phi$, by the mean apparent longitudinal field
strength $\left< B_{\rm app,los} \right>$, which relates to $\Phi$ by:
\begin{equation}
\label{eq_flux}
\Phi = \int_{\rm box} B_{\rm app,los} {\rm d}A = \left< B_{\rm app,los} \right> A_{\rm box},
\end{equation}
where $A_{\rm box}$ stands for the box area. $\left< B_{\rm app,los} \right>$ was chosen instead of $\Phi$
because the weak network boxes are larger than the strong network and plage ones. Thus, $\left< B_{\rm app,los}
\right>$ represents the degree of concentration of the magnetic flux, or ``magnetic activity'', in the local area
delimited by the box. Note that pores were included in the calculation of $\left< B_{\rm app,los} \right>$, as
they also contribute to the inhibition of convection. However, as our study focuses mainly on the contrast of
magnetic elements, pores and their immediate surroundings were removed from the subsequent contrast analysis
using the same method as in Paper I (the yellow contours in the top left panel of Fig. \ref{fig_boxes} outline
such pores).

In each box, the contrast at a pixel location ($x,y$) was defined relative to the mean continuum intensity
$\left< I_c \right>_{\rm ref,box}$ of all the pixels in the box having $B_{\rm app,los} < 100$ G, which
correspond to weakly magnetized areas harbouring relatively normal granulation (see Fig. \ref{fig_boxes}):
\begin{equation}
\label{eq_contrast}
{{\rm Contrast} (x,y)} = \frac{I_c(x,y) - \left< I_c \right>_{\rm ref,box}}{\left< I_c \right>_{\rm ref,box}}
\end{equation}
Using one contrast reference per box prevents the 5 min oscillations and possibly instrumental variations
\citep[e.g. varying amounts of defocus of the Hinode/SOT between scans,][]{Danilovic08} to introduce spurious
scatter in the contrast data.\footnote{This implicitely assumes that the intrinsic brightness of the weakly
magnetized areas $B_{\rm app,los} < 100$ G does not vary from one box to the other.}

Likewise, the LOS velocities $v_{\rm los}$ were calibrated independently for each box, since oscillations can
induce a \emph{global} up- or downflow of an entire box, while we are interested in quantifying the \emph{local} up- and
downflows within the box. We thus calibrated the velocities in each box such that the average of $v_{\rm los}$
over the areas where $B_{\rm app,los} < 100$ G is equal to the Doppler convective blueshift of the 630.2 line
minimum as derived from the FTS spectrum at the spectral resolution of Hinode (200 m s$^{-1}$).

\section{Results}
\label{sec_results}

\subsection{Contrast of magnetic elements and surrounding flows}
\label{sec_rescontrast}

\begin{figure}
\centering
\includegraphics[width=0.5\textwidth]{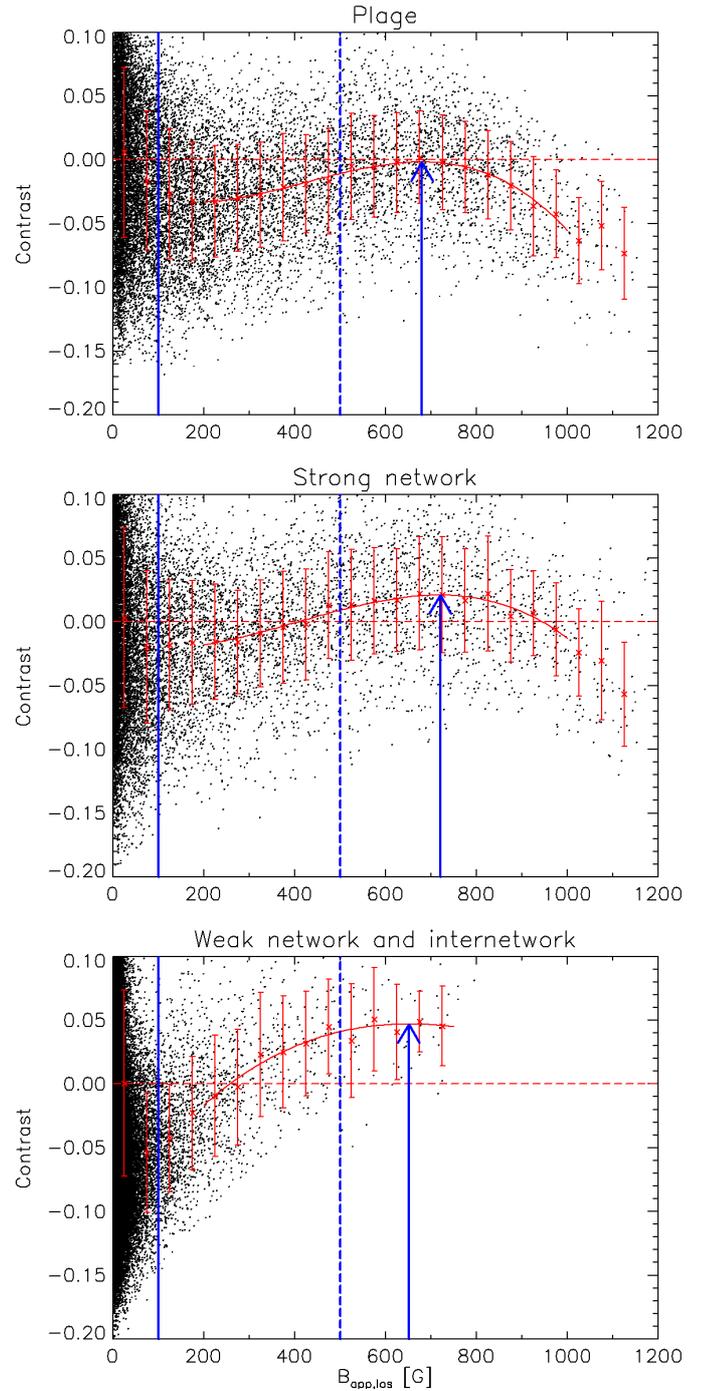}
\caption{Scatterplots of continuum contrast vs. apparent longitudinal field strength $B_{\rm app,los}$ for the
boxes shown in Fig. \ref{fig_boxes}, pores excluded. Red crosses: average values of the continuum contrast inside
$B_{\rm app,los}$-bins of 50 G width. The red error bars are the standard deviations inside each bin. Solid red
curves are third-order polynomial fits of the average values in the range 200 G $< B_{\rm app,los} < 1000$ G.
Blue arrows: indicate the peaks of the trend of the contrast vs. $B_{\rm app,los}$. The vertical solid blue lines
at $B_{\rm app,los} = 100$ G separate the ``weakly magnetized areas'' used as contrast and velocity reference and
the ``magnetic patches''. The dashed blue lines at $B_{\rm app,los} = 500$ G is the upper limit below which
pixels can be considered as not belonging to magnetic features.}
\label{fig_C_vs_B_boxes}
\end{figure}

In the first place we investigated how the contrast of magnetic elements varies with the amount of unsigned
magnetic flux contained in the boxes. Therefore we made for each box a scatterplot of the continuum contrast vs.
$B_{\rm app,los}$. We then proceeded as in Paper I, namely by averaging the pixel contrasts in $B_{\rm app,
los}$-bins of 50 G width, and fitting these average values by a 3$^{\rm rd}$-order polynomial (see Fig.
\ref{fig_C_vs_B_boxes}). In the following we shall simply refer to the maximum of the polynomial
trend as \emph{contrast peak}, which can be considered as a measure of the (average) contrast of the brightest
magnetic elements in the box. Figure \ref{fig_C_vs_B_boxes} presents the scatterplots of the
contrast vs. $B_{\rm app,los}$ for the three boxes presented in Fig.~\ref{fig_boxes}, along with their trends and
respective peaks (indicated by the arrows). Note how the trends vary between the three boxes: the peak increases
from plage to weak network while the range of $B_{\rm app,los}$ decreases. Together with the fact that the peak
contrast occurs at $B_{\rm app,los} \sim 700$ for all boxes, these observations are consistent with Paper I.
Hence, despite the relatively small size of the boxes, they contain enough statistics to accurately determine the
contrast peak. Note that for the weak network box, the trend is almost monotonically increasing. This is due to
the absence of dark features such as micropores, which are common in the strong network and plages. Such
quasi-monotonic trends were also obtained by \citet{Schnerr10} and \citet{Viticchie10}, who considered fields of
view containing QS network and internetwork, observed at SST and with IBIS, respectively.

\begin{figure*}
\centering
\includegraphics[width=\textwidth]{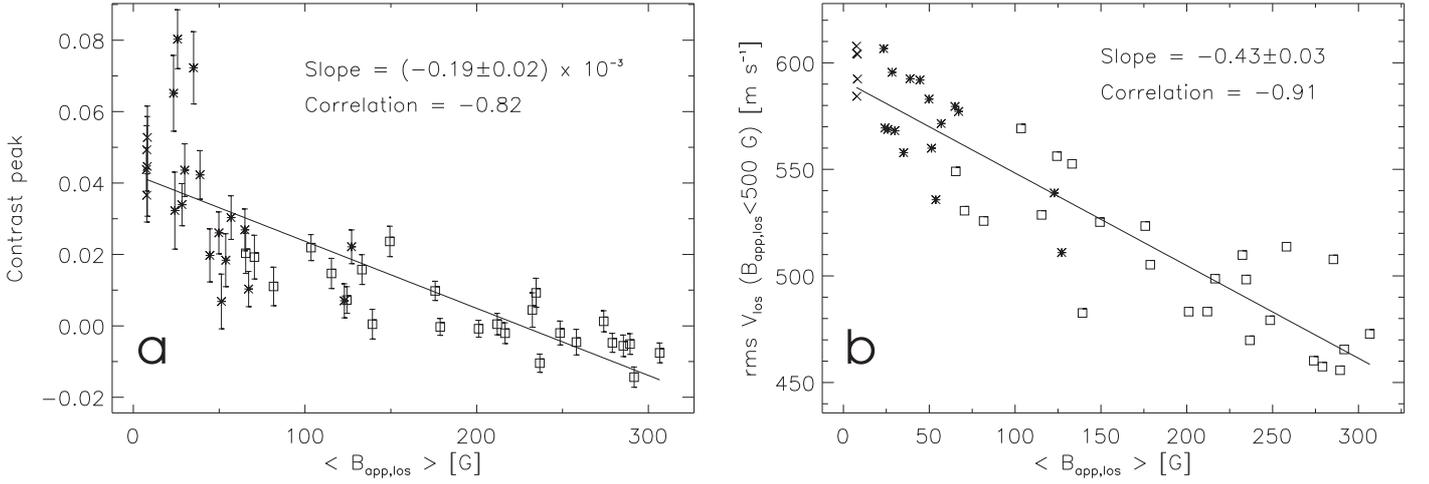}
\caption{a) Contrast peak vs. mean longitudinal field strength
$\left<B_{\rm app,los}\right>$ for the plage (``$\Box$''), strong network (``*'') and weak network boxes (``$\times$'').
b) RMS of the LOS velocity $v_{\rm los}$ outside strong magnetic features ($B_{\rm app,los} < 500$ G) vs.
$\left<B_{\rm app,los}\right>$ for the same boxes. Solid lines: linear regressions.}
\label{fig_CVlos_vs_avgB}
\end{figure*}

Figure \ref{fig_CVlos_vs_avgB}a displays the relation between the contrast peak and $\left< B_{\rm app,los}
\right>$ in the box. It can be seen that both quantities clearly anti-correlate, such that the magnetic elements
are darker in the more active boxes. The relationship is linear for $\left< B_{\rm app,los} \right> > 50$ G, but
it cannot be excluded that the contrast of magnetic elements in regions with $\left< B_{\rm app,los} \right> \leq
50$ G might have a significantly stronger dependence on $\left< B_{\rm app,los} \right>$ than given by the
regression line.

To test the idea that the convective energy transport surrounding the magnetic elements can affect their
contrasts (see Sect. \ref{sec_intro}), we also determined the rms of the line-of-sight (LOS) velocity $v_{\rm
los}$ (returned by the inversions) as a measure of the vigour of the vertical convective flows in the boxes.
Since we were only interested in characterizing the flows \emph{surrounding} the magnetic features, we restricted
the measurement of rms$(v_{\rm los})$ to the pixels where $B_{\rm app,los} < 500$ G.\footnote{It is already
well-known that convection is strongly suppressed \emph{within} the magnetic elements due to the kG strength of
the field, see, e.g., the review of \citet{Schuessler_rev92}.} As can be seen from Fig. \ref{fig_C_vs_B_boxes}
(dashed vertical line), this relatively arbitrary threshold value is sufficiently far from the peak of the
contrast vs. $B_{\rm app,los}$ trends (occuring at $B_{\rm app,los} \sim 700$ G) to consider that those pixels do
\emph{not} belong to those magnetic elements producing the peak, nor to the stronger and darker magnetic
features.\footnote{The same threshold was already adopted in Paper I.} The locations where $B_{\rm app,los} <
500$ G are delimited by green contours in Fig. \ref{fig_boxes} (middle and bottom panels). In fact, pixels with
$100 < B_{\rm app,los} < 500$ G correspond to abnormal granulation surrounding the magnetic features (see below).
Their magnetic signal probably comes from the expanding canopies of the magnetic features
\citep[e.g.][]{Solanki90} as well as a ``bloom'' of scattered polarization signal \citep{Lites02}. However, it
cannot be excluded that such pixels also contain magnetic elements with small filling factors. Figure
\ref{fig_CVlos_vs_avgB}b then reveals that the rms$(v_{\rm los})$ of the pixels with $B_{\rm app,los} < 500$ G
linearly decreases with the $\left< B_{\rm app,los}\right>$ in the boxes. Note that the slope of the regression
and correlation coefficient depend only weakly on the chosen upper limit of $B_{\rm app,los}$: if we chose a
threshold of 300 G instead, they would vary by less than 3\%. Note that, as a consequence of the
anti-correlations of both the contrast peak and rms$(v_{\rm los})$ with $\left< B_{\rm app,los}\right>$, the
contrast peak correlates with rms$(v_{\rm los})$ (with a coeff. 0.79, plot not shown here). The decrease of both
the contrast peak and rms$(v_{\rm los})$ with $\left< B_{\rm app,los} \right>$ supports the idea that as the flux
gets more concentrated in the boxes, the contrast of the magnetic elements is reduced as a result of a less
efficient convection.

\subsection{Magnetic patches}


A closer look at the boxes gave us more insight as to how the embedded magnetic flux in the box can hamper the
convection and affect the contrast of magnetic elements. Inside plages and strong network, most of the flux is
contained in patches that have rather sharp boundaries \citep{Title92}, also referred to as ``magnetic islands''
by \citet{Ishikawa07} and ``extended contiguous regions of strong field'' by \citet{Narayan10}. In our case these
areas can be relatively well enclosed by contours of $B_{\rm app,los} > 100$ G (see Fig. \ref{fig_boxes}). These
patches appear composed by clusters of small magnetic elements, micropores, pores and disturbed granulation (see
also Sect. \ref{sec_discflows}).

They contain between 60 and 95 \% of the total unsigned flux of the box, this fraction increasing rapidly with
$\left< B_{\rm app,los} \right>$ as the patches cover a larger part of the box (see below). As can be seen from
Fig. \ref{fig_Phi_fA_vs_avgB}a, the total unsigned flux $\Phi$ (see Eq. \ref{eq_flux}) contained in the box
almost perfectly scales (correlation coeff. $> 0.995$) with the total flux carried by the patches. Only the
strong network and plage boxes have been represented because they have the same areas (hence their fluxes can be
directly compared). Note that from the weak network to the plage boxes, the fractional area of the box covered by
the patches varies by about one order of magnitude (from 0.05 to 0.55), whereas the average value of $B_{\rm
app,los}$ computed over the \emph{patches} varies only between about 300 and 500 G (see absissa of Fig.
\ref{fig_rmsVlos_vs_avgB_patches}b).\footnote{The mean $B_{\rm app,los}$ of the patches naturally slightly
increases as they become larger, since there is more flux available for the formation of stronger and larger
magnetic features.} Hence, the amount of unsigned flux contained in a box, and thereby $\left< B_{\rm
app,los}\right>$, is mainly determined by the fraction of the box occupied by these patches, as illustrated in
Fig. \ref{fig_Phi_fA_vs_avgB}b (with a correlation of 0.99).
Since they carry most of the magnetic flux, we can attribute to the patches the degree of
inhibition of vertical convection and the decrease of the contrast of the magnetic elements (see Sect.
\ref{sec_discflows}). To show this, we will separate the contributions from the patches from their weakly
magnetized surroundings ($B_{\rm app,los} < 100$ G).

We shall first differentiate the state of convection inside and ouside the patches.
Therefore, we evaluated rms($v_{\rm los}$) and $\left< B_{\rm app,los} \right>$ separately in the areas outside
($B_{\rm app,los} < 100$ G) and inside the patches ($B_{\rm app,los} > 100$ G). For the latter, only the abnormal
granulation within the patches was considered (with an upper limit of 500 G, cf. Sect. \ref{sec_rescontrast}),
while magnetic elements and pores were included in $\left< B_{\rm app,los} \right>$ (as both contribute to the
inhibition of convection in their surroundings). The variation of rms($v_{\rm los}$) as a function of $\left<
B_{\rm app,los} \right>$ in the areas inside and outside magnetic patches is presented in Fig.
\ref{fig_rmsVlos_vs_avgB_patches}a and Fig. \ref{fig_rmsVlos_vs_avgB_patches}b, respectively. Comparing the two
Figs., we see that the rms($v_{\rm los}$) in the patches is on average lower by about 110 m s$^{-1}$,
corresponding to a relative reduction of $\sim 21$\% (the average values of rms($v_{\rm los}$) are 527 and 417 m
s$^{-1}$ outside and inside the patches, respectively).
That the $v_{\rm los}$ amplitudes are reduced within the patches is also visible in the lower panels
of Fig. \ref{fig_boxes}, especially for the plage boxes. Note that the correlation in Fig.
\ref{fig_rmsVlos_vs_avgB_patches}b is rather weak, mainly because $\left< B_{\rm app,los} \right>$ in the patch
varies only between 300 and 550 G (relative variation of 45\%). Also, the number of pixels satisfying $100 <
B_{\rm app,los} < 500$ G is relatively low, especially in the network boxes (these pixels cover only between 5
and 30 \% of the box area).
In contrast, outside the patches $\left< B_{\rm app,los} \right>$ varies by a factor 4, from 10 to 40 G,
therefore the rms($v_{\rm los}$) exhibits a strong correlation ($-0.91$) with $\left< B_{\rm app,los} \right>$ in
these areas. As we shall see below (last paragraph and Fig. \ref{fig_avgCref_vs_avgB}), however, the reduction of
rms($v_{\rm los}$) with $\left< B_{\rm app,los} \right>$ in these weakly-magnetized areas does not cause
a reduction of their average brightness $\left< I_c \right>_{\rm ref,box}$. The choice of $\left< I_c
\right>_{\rm ref,box}$ as contrast reference is thus also justified by its robustness with respect to variations
of $\left< B_{\rm app,los} \right>$.

Next, we inquired whether the mean field strength in the boxes also affects the averaged contrast of the magnetic
patches (outside the patches the contrast is equal to 0 by definition). Figure \ref{fig_avgC_vs_avgB_patches}a
displays the variation of the contrast averaged over the patches (pores excluded) as a function of $\left< B_{\rm
app,los} \right>$ for all the boxes. Note that the weak network boxes were not included here because their
patches are too small for proper statistics. Also, by our definition of contrast reference $\left< I_c
\right>_{\rm ref,box}$, the average contrast of the patches is equivalent to their average brightness relative
to their weakly magnetized surroundings. As can be seen, the patches are always darker than their surrounds
(except for 3 network boxes due to their bright magnetic elements), and become even darker as $\left<
B_{\rm app,los} \right>$ increases and the patches grow bigger (i.e. occupy a larger fraction of the boxes cf.
Fig. \ref{fig_Phi_fA_vs_avgB}b). Since the magnetic patches contain both magnetic elements and abnormal
granulation (pores being excluded from the analysis), we can further separate these two contributions. When
averaging the contrast over the magnetic patches while at the same time excluding magnetic elements (i.e.
considering only pixels with $100 < B_{\rm app,los} < 500$ G) as in Fig. \ref{fig_avgC_vs_avgB_patches}b, we
notice that the contrast values are now all negative and lower than in Fig. \ref{fig_avgC_vs_avgB_patches}a,
particularly for the QS network boxes from which we removed the bright magnetic elements (which are brighter than
in plage boxes cf. Fig. \ref{fig_CVlos_vs_avgB}a). This causes the slope of the regression line to decrease by a
relative difference of 34\%. We can thus conclude that about one third of the darkening of the patches with
increasing $\left< B_{\rm app,los} \right>$ can be attributed to the decreasing brightness of the magnetic
elements, while about two thirds is due to a darkening of the abnormal granulation. This is understandable as the
magnetic elements cover only between 10 and 40\% of the area of the patches (with increasing fraction from
network to plage boxes). In his 3D MHD simulations, \citet{Voegler05b} similarly observed that both the
decreasing intensity of the magnetic features and of the granulation were contributing to the darkening of the
simulation snapshots (for $\left<B\right>$ greater than 200 G). Note that the correlations visible in Fig.
\ref{fig_avgC_vs_avgB_patches}a,b persist if plotting against the mean $B_{\rm app,los}$ of the patch itself
(although the correlation are reduced to $-0.56$ and $-0.37$, respectively, but not shown here).

Finally, from the observation that the patches have negative contrast combined with the fact that they occupy a
larger fractional area of the boxes as $\left< B_{\rm app,los} \right>$ increases, one can deduce that the
contrast averaged over the \emph{entire} boxes (pores excluded) will also exhibit a decreasing trend with $\left<
B_{\rm app,los} \right>$, which is confirmed in Fig. \ref{fig_avgC_vs_avgB}. Note that the correlation
coefficient ($-0.89$) has improved compared to the correlation between the average contrast of the patches and
$\left< B_{\rm app,los} \right>$ ($-0.71$, cf. Fig. \ref{fig_avgC_vs_avgB_patches}a). This is because the
decrease of the box-averaged contrast with $\left< B_{\rm app,los} \right>$ is mainly dictated by the rapid
increase of the fractional area of the patches with $\left< B_{\rm app,los} \right>$ (from 0.05 to 0.55, cf. Fig.
\ref{fig_Phi_fA_vs_avgB}b), while the darkening of the patches with $\left< B_{\rm app,los} \right>$ contributes
only marginally (it varies by about 1.5\% only). In spite of the good correlation, the average box contrast
decreases by only $\sim 1\%$, as the average includes the areas of the box outside the patches, where the
contrast is equal to 0 by definition. Hence, an area of photosphere in the 630 nm continuum at solar disk center
containing strong network or plage appears darker than a quiet one, and its darkness increases with increasing
magnetic flux.

To validate our contrast results, we checked that they were not biased by any systematic increase or decrease of
the intrinsic brightness of the areas taken as contrast reference in each box (pixels with $B_{\rm app,los}<100$
G, see Sect. \ref{sec_boxes}). A priori, the brightness of these areas could be affected by partial disturbances
of convection (see the decrease of the rms($v_{\rm los}$) in Fig. \ref{fig_rmsVlos_vs_avgB_patches}a), enhanced
radiative escape through a magnetized atmosphere, cooling by horizontal heat gradient toward the magnetic patches
\citep{Deinzer84} and acoustic oscillations. To see if these effects induce any systematic trend, we compared in
Fig. \ref{fig_avgCref_vs_avgB} the average \emph{intensity} values $\left< I_c \right>_{\rm ref,box}$ (in
``instrumental data numbers'', i.e. units obtained after calibration with \verb"sp_prep", see Sect.
\ref{sec_scans_inversions}) of the contrast reference areas in the boxes as a function of $\left< B_{\rm app,los}
\right>$. As the weak correlation demonstrates, there is no systematic trend of $\left< I_c \right>_{\rm
ref,box}$ with $\left< B_{\rm app,los} \right>$, which validates the appropriate threshold of 100 G for the
contrast reference. As indicated by the right-hand side y-axis of the plot (where the values are normalized by
the mean of all the $\left< I_c \right>_{\rm ref,box}$), the linear regression varies by less than 0.5\% over the
range of $\left< B_{\rm app,los} \right>$. Hence, variations of the contrast reference from box to box cannot
explain the decrease of 6\% of the contrast peak (from 0.04 to $-0.02$, see Fig. \ref{fig_CVlos_vs_avgB}a) and
the decrease of $\sim 1\%$ of the box-averaged contrast (Fig. \ref{fig_avgC_vs_avgB}) with $\left< B_{\rm
app,los} \right>$. Note that the random fluctuations of $\left< I_c \right>_{\rm ref,box}$ can be attributed to
oscillations and possibly to instrumental variations (cf. Sect. \ref{sec_boxes}). Altogether, they produce a
rms scatter of 1.3\%, much less than the scatter of the data points in Fig. \ref{fig_CVlos_vs_avgB}a and of
similar order than the scatter in Fig. \ref{fig_avgC_vs_avgB}. The relative stability of the contrast references
enables us to interpret our contrasts as an indicator of the \emph{intrinsic} brightness.

\begin{figure*}
\centering
\includegraphics[width=\textwidth]{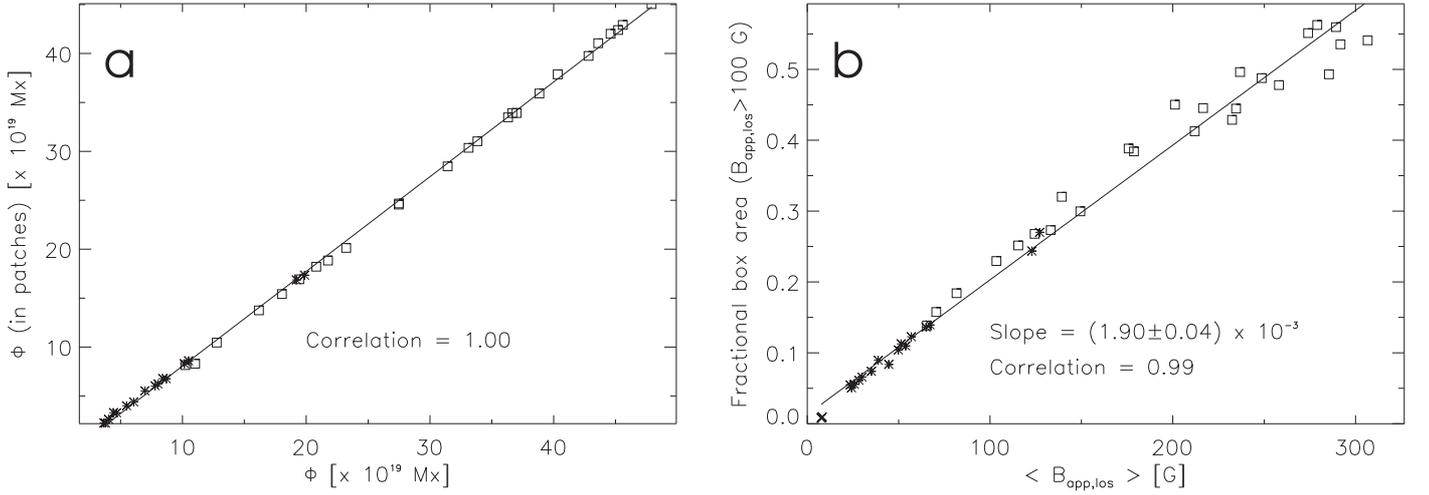}
\caption{a) Total unsigned magnetic flux carried by the magnetic patches ($B_{\rm app,los} >$ 100 G) vs. total
flux in the boxes for the strong network (``*'') and plage (``$\Box$'') boxes.
b) Fraction of the box's area occupied by the magnetic patches vs. mean longitudinal
field strength $\left<B_{\rm app,los}\right>$, for all the boxes including weak network (``$\times$'').}
\label{fig_Phi_fA_vs_avgB}
\end{figure*}

\begin{figure*}
\centering
\includegraphics[width=\textwidth]{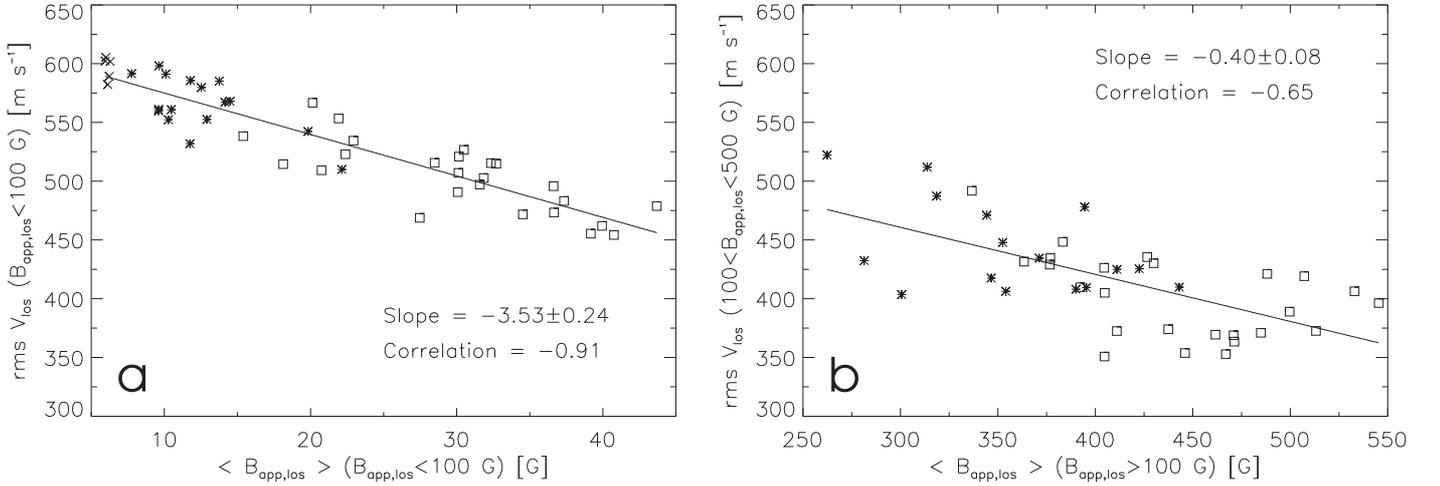}
\caption{a) RMS of the LOS velocity $v_{\rm los}$ vs.
$\left<B_{\rm app,los}\right>$ outside the magnetic patches ($B_{\rm app,los} < 100$ G).
b) RMS of the LOS velocity $v_{\rm los}$ vs.
$\left<B_{\rm app,los}\right>$ inside the magnetic patches ($B_{\rm app,los} > 100$ G), but excluding magnetic
features from the computation of rms($v_{\rm los}$) by restricting it to pixels with $B_{\rm app,los} < 500$ G.
}
\label{fig_rmsVlos_vs_avgB_patches}
\end{figure*}

\begin{figure*}
\centering
\includegraphics[width=\textwidth]{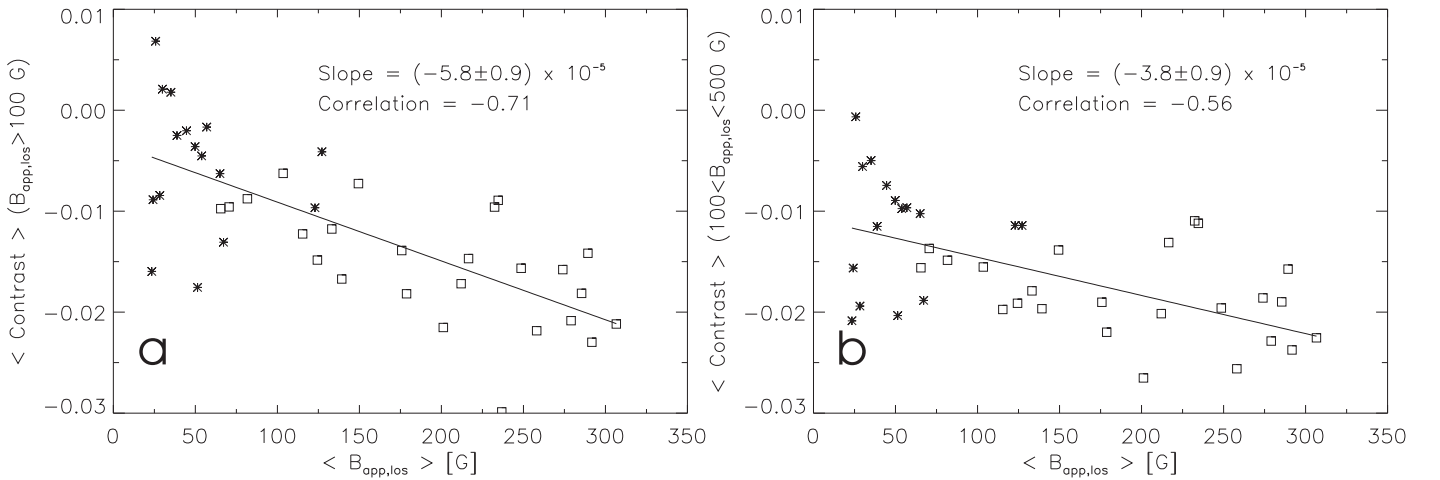}
\caption{a) Average contrast over the magnetic patches ($B_{\rm app,los} > 100$ G, pores removed) vs. mean
longitudinal field strength in the boxes $\left<B_{\rm app,los}\right>$, for the strong network (``*'') and plage
(``$\Box$'') boxes.
b) Same but excluding magnetic features from the average contrast ($100 < B_{\rm app,los} < 500$ G).}
\label{fig_avgC_vs_avgB_patches}
\end{figure*}

\begin{figure}
\centering
\includegraphics[width=\columnwidth]{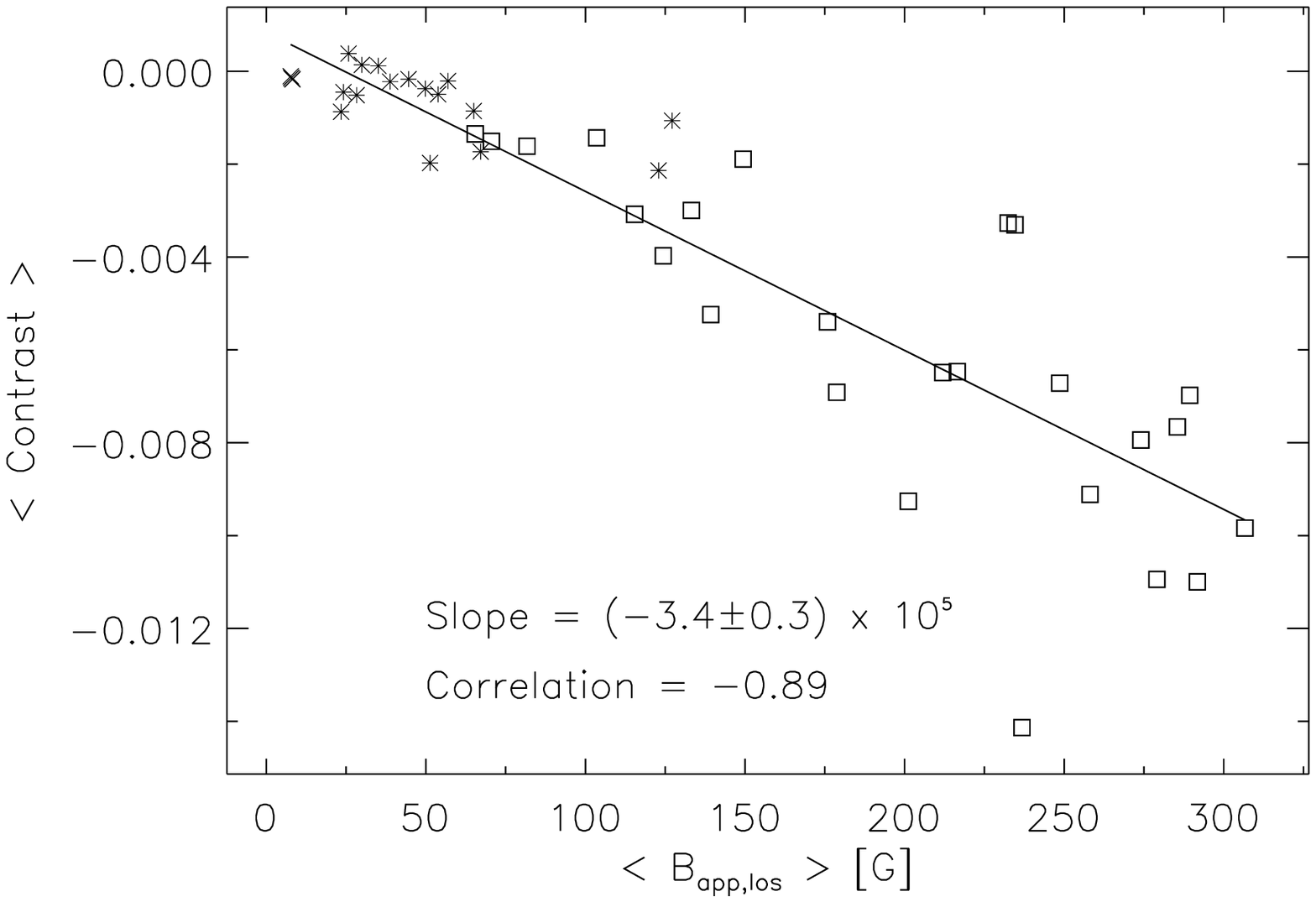}
\caption{Average of the contrast over the whole box (pores removed)
vs. $\left<B_{\rm app,los}\right>$.}
\label{fig_avgC_vs_avgB}
\end{figure}

\begin{figure}
\centering
\includegraphics[width=\columnwidth]{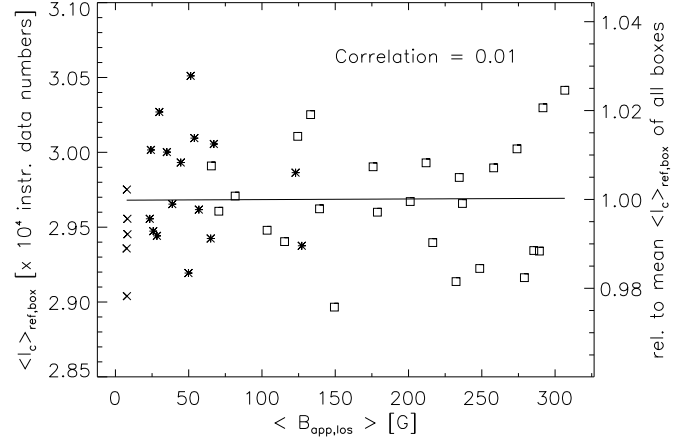}
\caption{
Variation of the mean intensity $\left< I_c \right>_{\rm ref, FOV}$ of the contrast reference areas in each box ($B_{\rm app,los} < 100$ G), as a function of $\left<B_{\rm app,los}\right>$ in the box.
$\left< I_c \right>_{\rm ref, FOV}$ is in instrumental data number.
The solid line is a linear regression.
The right-hand side y axis gives the values relative to the mean value of all boxes.}
\label{fig_avgCref_vs_avgB}
\end{figure}


\section{Discussion}
\label{sec_disc}

\subsection{Dependence of the contrasts on the mean field strength}
\label{sec_disccontrast}

Our box test showed that the contrast of the magnetic elements at disk center decreases continuously as the
magnetic flux in the local environment defined by the box increases (see Fig. \ref{fig_CVlos_vs_avgB}a).
In particular, the absence of a gap between the network and the ARs implies that the physical mechanisms
responsible for the brightness of the magnetic elements are primarily sensitive to the local mean field strength rather
than to the elements' location within network or plage.
According to the above, we expect the largest contrasts for nearly
isolated bright points in weak network and internetwork regions.

At first sight, this result seems consistent with the conclusions of earlier spectra-based studies
\citep[e.g.][]{Solanki84, Zayer90, Solanki92}, which found on average darker magnetic features in ARs than in the
QS. Although the studies based on Stokes profiles collect information originating solely from the magnetic
features, their low spatial resolution \citep[up to $10\arsec$ in the case of FTS
observations,][]{Stenflo84, Stenflo85} implies that the temperature information of magnetic elements may be mixed
with that of micropores and pores also present in the resolution element. Hence, for these spectra-based
studies the lower brightness in ARs can conceivably be due, in part or as a whole, to the presence of larger and
darker magnetic features. An interpretation in terms of average magnetic feature size was also favoured by
\citet{Grossman94}, on the basis of radiation-MHD models of magnetic features with different cross-sectional
areas. The same conclusions were reached by \citet{Ortiz02} using MDI data (angular resolution
$4\arsec$).
However, we believe that a variation of the size of the magnetic elements alone cannot explain our result,
because we specifically looked at the peak of the contrast vs. $B_{\rm app,los}$ trend in each box. As shown in
Paper I (Fig. 3), the magnetic elements contributing to this peak share similar magnetic filling factors, which
at Hinode's resolution can be assumed to scale with the cross-sectional area of the unresolved magnetic features.

This being said, one cannot exclude the possibility that the decrease of the contrast peak with $\left<B_{\rm
app,los}\right>$ is indirectly influenced by the increasing number of larger features including micropores. Such
features that could not be removed by our pore removal procedure are susceptible to have pixels with similar
$B_{\rm app,los}$ as the bright magnetic elements, especially towards their edges (see Paper I). The mixing of
such relatively dark pixels with bright magnetic element pixels in the same $B_{\rm app,los}$-bin could then
lower the contrast values. Telescope diffraction is also likely to play a role, as magnetic elements are often
located in the direct vicinity of larger and darker flux concentrations (see Paper I Fig. 1). However, both these
effects should lead to a larger dispersion in the scatterplots around the contrast peak (Fig.
\ref{fig_C_vs_B_boxes}) for the AR boxes, which we do not observe.

\begin{figure*}
\centering
\includegraphics[width=\textwidth]{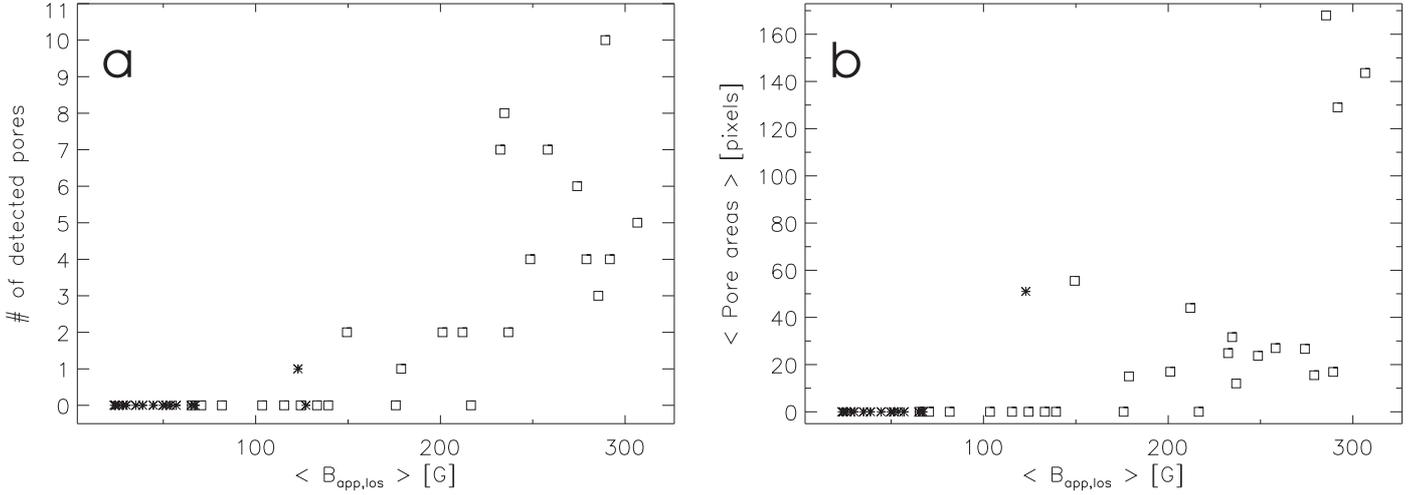}
\caption{a) Number of pores in the strong network (``*'') and plage (``$\Box$'') boxes as a function of
$\left<B_{\rm app,los}\right>$. b) Mean area of the pores (in number of pixels) in each box vs. $\left<B_{\rm
app,los}\right>$.}
\label{fig_na_cores}
\end{figure*}

To have an idea on the magnitude of these possible effects, it is instructive to look at how the number and area
of larger magnetic features vary with the mean field in the boxes. Pores constitute a suitable target for that
because they are generally sufficiently separated, so that their areas can be measured individually and are large
enough to be easily segmented, even at Hinode's resolution. We thus quantified the number of pores and mean area
of a pore (i.e. total area covered by the pores divided by the number of pores) in each of our
strong network and plage boxes. To do so, we restricted the pores to their dark ``cores'' defined as the set of
pixels where the contrast $< -0.15$ and $B_{\rm app,los} > 900$ G (cf. Paper I).
As can be seen from Fig. \ref{fig_na_cores}a and Fig. \ref{fig_na_cores}b, when $\left<B_{\rm app,los}\right>$
reaches 100-150 G, both the number and mean area (number of pixels per pore) increase rapidly with $\left<B_{\rm
app,los}\right>$. If this is representative of the increasing number of larger-than-magnetic element features
(not only pores), then it means that the effect of larger and darker features on the contrast of magnetic
elements would only become significant for $\left<B_{\rm app,los}\right> > 150$ G, i.e. for the plage boxes.

Of direct relevance for the solar irradiance contributions near disk center is the decrease of the
contrast averaged over the whole box as the magnetic flux increases. This result confirms and extends the
finding of \citet{Title92} that the mean continuum contrast (at 676.8 nm) of the single plage they observed (at
$\mu = 0.97$) was lower than the QS value. Although we only probed one wavelength, our result indicates a
reduced vertical energy output with larger $\left<B_{\rm app,los}\right>$, as already predicted by the 3D MHD
simulations of \citet{Voegler05b}. Note that the curve of mean bolometric intensity vs. mean field strength
$\left<B\right>$ of \citet{Voegler05b} does not drop as linearly as ours and first exhibits an increase for
$\left<B\right> < 100$ G. In particular, the simulation boxes of \citet{Voegler05b} yield mean bolometric
intensities dropping below the QS level only for $\left<B\right>$ greater than 200 G, whereas our
box-averaged contrasts are negative even for mean longitudinal field strengths lower than 200 G (see Fig.
\ref{fig_avgC_vs_avgB}). These discrepancies can be attributed to different factors: an obvious one is that our
measurements are monochromatic, so that they provide only a rough guide to the total energy loss, in particular
that emitted at higher layers of the photosphere (e.g. in spectral lines). Another important factor is that the
magnetic field in the simulations distributes itself more or less homogeneously in the intergranular lanes of the
simulation box (i.e. in some ways like an infinitely large patch), whereas in the observations the magnetic
elements tend to be clumped in patches, which are mainly responsible for the decrease in intensity and have a
considerably higher average $B_{\rm app,los}$ than the whole box. The cancellation of mixed polarities in our
pixels and the fact that we only measure the longitudinal component of the field may also contribute. A visual
comparison with the snapshots of \citet{Voegler05b} also gives the impression that the magnetic patches within
our boxes of strong network and plage (where we find negative average contrasts) rather resemble their box
with $\left<B\right> = 400$ G, for which the disk center bolometric intensity is lower than the QS value.


\subsection{Inhibition of convective flows}
\label{sec_discflows}

The decrease of the contrast of the magnetic elements (Fig. \ref{fig_CVlos_vs_avgB}a), the contrast of the
patches (Fig. \ref{fig_avgC_vs_avgB_patches}) and the box-averaged contrast (Fig. \ref{fig_avgC_vs_avgB}) with
$\left<B_{\rm app,los}\right>$ could partly be caused by the inhibition of convection in the presence of magnetic
fields. That the close packing of magnetic elements in plage could lead to a significant disturbance of the
surrounding convective energy transport and thereby to lower contrasts was already proposed by
\citet{Knoelker88b, Knoelker88}. This is supported by observational results from spectral lines indicating a
weaker convective transport in regions with larger filling factors \citep{Livingston82, Cavallini85,
Immerschmitt87, Brandt90}. Based on Hinode/SP data, \citet{Morinaga08} recently gave evidence of the local
action of the magnetic field on the flows, by observing that the vertical convection was more suppressed in
locations (pixels) of larger apparent field strength.
This is now validated on the larger scale of our boxes by showing that, as the amount of magnetic flux increases
in the box, the convective flows surrounding the magnetic elements are progressively more suppressed (Fig.
\ref{fig_CVlos_vs_avgB}b).

Note that the linear relationship between rms($v_{\rm los}$) outside the magnetic elements and $\left< B_{\rm app,los}
\right>$ is \emph{independent of the box size}. We indeed repeated the analysis using boxes with side length
varying from $10\arsec$ to $70\arsec$ centered at the same locations, and found that the slope of the linear
regression did not vary by more than $5\%$. This intuitively means that the convective flows are affected by the
mean magnetic flux \emph{density} and not by the total amount of flux.

Magnetic elements are not homogeneously distributed in the box, but located inside magnetic patches, which have
been previously discussed by \citet{Ishikawa07} and \citet{Narayan10}. Like us (see Fig.
\ref{fig_rmsVlos_vs_avgB_patches}), they found that the dispersion of the vertical velocities is significantly
reduced in the patches compared to their surroundings.
In the boxes presented in Fig.
\ref{fig_boxes}, one can see that in the continuum (upper panels) the patches have a diffuse fluid-like
appearance. The granulation therein can hardly be recognized while the vertical velocities (lower panels) appear
significantly reduced compared to the surroundings. Both these effects are probably due to the high density of
magnetic features in the patches, which locally hampers the convective motions (Fig.
\ref{fig_rmsVlos_vs_avgB_patches}b) causing the spatial scale of convection to become smaller (``abnormal'').
Telescope diffraction also contributes to the diffuse appearance of the patches. The combination of
both effects could lead us to underestimate the rms($v_{\rm los}$) inside our patches, inasmuch as the velocity
field might be partly under-resolved. With the help of SST CRISP data at $0\dotarsec15$ resolution,
\citet{Narayan10} were indeed able to identify a small-scale convection pattern within their patches and
estimated its spatial scale to be $\sim0\dotarsec3$, i.e. similar to Hinode's resolution. Hence, as the amount of
flux in the box increases, the reduced vertical
velocities and reduced convection scale in the patches have the statistical consequence of decreasing the
rms($v_{\rm los}$) in the box.

The magnetically-disturbed convection \emph{within} the patches (Fig. \ref{fig_rmsVlos_vs_avgB_patches}b) is also
likely to be the cause of the reduced brightness of the abnormal granulation (Fig.
\ref{fig_avgC_vs_avgB_patches}b), just as deduced by \citet{Voegler05b} from his simulations. The partial
inhibition of convective motions and the reduced spatial scale of convection limit the convective heat upflow and
the efficiency of the convective energy transport, which tends to cool the patch. Hence the patch becomes ever
cooler and darker as its mean field strength and area increase, because both contribute to inhibit the convection. The
inhibition of convection is driven by the patch size to a certain extent, since for a very small patch no granule
is completely surrounded by magnetic features, while for a larger patch an increasing number of granules is. We
speculate that the larger mean field strength of the patch is mostly due to an increased average size of magnetic
features, which are then better resolved - i.e. have a larger magnetic filling factor - and due to the formation
of somewhat stronger magnetic features. Since larger features tend to be darker, this is also likely to
contribute to the darker appearance of the patch.

As magnetic elements are located inside the patches, their brightness should be affected by the local amount of
heat available in the patch ``reservoir''. This would explain the decrease of the brightness of the magnetic
elements from network to plage as the patches become larger and at the same time aquire a larger mean field. This
interpretation is consistent with the finding of \citet{Ishikawa07} that G-band bright points tend to lie
preferentially close to the boundary of the magnetic patches, as the further inside they get, the farther they
are from the heat source of normal granulation. Note that the fact that the magnetic patches appear darker than
normal granulation combined with telescope diffraction could also play a role when comparing the observed
contrasts of magnetic elements in ARs, which are mostly located within large patches, with the contrast of QS
bright points, which are often isolated and surrounded by normal granulation \citep[see][and Paper
I]{Lawrence93}. The limited resolution of the telescope that blends the signals of the BPs could thus have a
stronger darkening effect in ARs than in QS, although isolated QS BPs are also susceptible of being smeared
out in the intergranular lanes \citep[][]{Title96}. However, we believe this is not the main effect causing the decrease of the
contrast peak from the QS network to the AR plage boxes, as the mean contrast of the abnormal granulation in the
patches (i.e. with 100 G $< \left<B_{\rm app,los}\right> <$ 500 G) varies by only $\sim 1$\% over the range of
$\left<B_{\rm app,los}\right>$ (Fig. \ref{fig_avgC_vs_avgB_patches}b), whereas the contrast peak decreases by
$\sim 6$\% (Fig. \ref{fig_CVlos_vs_avgB}a).


\section{Summary and conclusions}
\label{sec_concl}

Previous studies found that at disk center, the continuum intensity contrast in the QS network
reached higher values than in ARs \citep{Title89, Topka92, Lawrence93} and \citep{Topka97}. This
was confirmed by the first paper of this series \citep[][Paper I]{Kobel11a} using Hinode/SP data, where we
established that the relation between continuum brightness (at 630.2 nm) and longitudinal field strength
$B_{\rm app,los}$ (retrieved by Milne-Eddington inversions) exhibits a higher peak in the QS network than in ARs.
Since the magnetic elements producing the peak share similar magnetic filling factors, we argued that this
brightness difference between the QS and ARs is unlikely to be explained solely by a difference in the size of
the brightest magnetic elements (although the presence of larger features in ARs certainly plays a role in
determining their brightness more generally, see below).

In this paper, we tested whether the contrast of the magnetic elements at disk center could be affected
by an altered convective transport due to the local concentration of magnetic flux. With this aim, we extracted
from Hinode/SP scans a series of ``boxes'' containing different amounts of magnetic flux, covering areas from AR
plages to weak QS network. We found that the contrast of the brightest magnetic elements (i.e. the peak of the
contrast vs. $B_{\rm app,los}$) continuously decreases with the mean longitudinal field strength $\left< B_{\rm
app,los} \right>$ in the boxes (Fig. \ref{fig_CVlos_vs_avgB}a). Since the $\left< B_{\rm app,los} \right>$ is
generally larger in ARs than in the QS, this can be taken as a generalization of the earlier finding that
magnetic elements in the network are brighter than in ARs.

Hence, as the amount of flux (and thereby $\left< B_{\rm app,los} \right>$) increases in a local area, the
observed contrast of the magnetic elements associated to the contrast peak decreases due to a combination of two
effects: (i) The known presence of larger (and thus darker) magnetic features, even if pores are
removed. These can either provide pixels. Such larger magnetic features reduce the contrast
of nearby magnetic elements through the telescope point spread function. Yet this effect is likely to play
only a minor role in the decrease of the magnetic element contrasts with $\left< B_{\rm app,los} \right>$ (see
discussion in Sect. \ref{sec_disccontrast}).
(ii) The disturbance of convective energy transport in the surrounding of the magnetic features. This is
supported by the anti-correlation between the fluctuation of the longitudinal velocities outside the magnetic
features and $\left< B_{\rm app,los} \right>$ (Fig. \ref{fig_CVlos_vs_avgB}b) and the lower brightness of the
abnormal granulation surrounding magnetic elements compared to weakly magnetized areas (Fig.
\ref{fig_avgC_vs_avgB_patches}b).

A closer inspection of the strong network and plage boxes revealed that the magnetic flux in these boxes is
mostly carried by ``magnetic patches'' that are well delimited by $B_{\rm app,los} > 100$ G (Fig.
\ref{fig_Phi_fA_vs_avgB}a). The flux density in these patches is such that the vertical convective velocities
inside them (around magnetic features) are reduced significantly relative to their surroundings (the rms($v_{\rm
los}$) is about 100 m s$^{-1}$ lower), this reduction being more important as the mean field of the patches
increases (cf. Fig. \ref{fig_rmsVlos_vs_avgB_patches}b). These reduced velocities probably limit the heat
upflow and consequently lower the brightness of the granulation in the patches compared to the granulation in the
weakly magnetized areas (Fig. \ref{fig_avgC_vs_avgB_patches}b). The reduced spatial scale of the (abnormal)
granulation inside the patches probably also contributes to decreasing the efficiency of the convective energy
transport. Even with their magnetic elements, the patches are generally darker than their surroundings (Fig.
\ref{fig_avgC_vs_avgB_patches}a), and become all the more dark when $\left< B_{\rm app,los} \right>$ increases in
the box and the patches occupy a larger part of it (Fig. \ref{fig_Phi_fA_vs_avgB}b). We interpret the patches as
having the following influence on the brightness of the magnetic elements: \emph{the more extended the magnetic
patch is (independently of the box size) and the larger its mean field, the larger is the inhibition of
convection, resulting in cooler granules and smaller heat influx to the ``hot walls'' of the magnetic elements
which become cooler and less bright (at least at disk center)}.

Since the patches are darker than their surrounding and occupy a larger part of the box as $\left< B_{\rm
app,los} \right>$ increases, the average brightness of a box at disk center also decreases with increasing $\left< B_{\rm
app,los} \right>$ (by about 1.2\% from weak network to plage boxes). Previous 3D MHD simulations of
\citet{Voegler05b} already predicted a reduction of the vertical bolometric intensity of boxes with mean field strength
$\left< B\right> > 200$ G. The qualitative agreement between our results and these simulations (for $\left<
B\right> > 200$ G) supports their use to model the different component atmospheres used in irradiance
reconstructions \citep[see, e.g.,][]{Unruh09}.

Up to now, the effect of the inhibition of convection on the photospheric contrast has not been taken into
account in irradiance reconstruction models. These models only use the photometric area of magnetic features
\citep{Penza03, Fontenla05} or their ``filling factor'' within a pixel \citep[in the case of SATIRE, see][for
reviews]{Solanki05, Krivova10} as the time-dependent variables modulating the intensity output given by
semi-empirical 1D model atmospheres. Our results suggest that the reconstructions could be possibly enhanced by
modulating the emitted intensity (e.g. for a given magnetic filling factor) according to the surrounding magnetic
flux in a local (e.g. in a $20\arsec \times 20\arsec$) area.
However, the finding that the inhibition of convection in more active areas affects the brightness of magnetic
elements has been tested here for the disk center only. In the next article of this series, we shall investigate
this effect away from the disk center by comparing center-to-limb variation of the continuum contrast of magnetic
elements in the QS and in ARs separately.

\begin{acknowledgements}
We gratefully thank M. Sch\"{u}ssler, D. R\"{o}rbein and the Solar Lower Atmosphere and Magnetism (SLAM) group at
the Max-Planck Institut f\"{u}r Sonnensystemforschung, as well as J. de la Cruz Rodr{\'{\i}}guez, Y. Unruh, W.
Ball, B. Viticchi\'{e}, N. Krivova and J. Sanchez-Almeida for their interest and fruitful discussions about this
work. This work has been partially supported by WCU grant No. R31-10016 funded by the Korean Ministry of
Education Science and Technology. Hinode is a Japanese mission developed and launched by ISAS/JAXA, collaborating
with NAOJ as a domestic partner, NASA and STFC (UK) as international partners. Scientific operation of the Hinode
mission is conducted by the Hinode science team organized at ISAS/JAXA. This team mainly consists of scientists
from institutes in the partner countries. Support for the post-launch operation is provided by JAXA and NAOJ
(Japan), STFC (U.K.), NASA, ESA, and NSC (Norway). This work has also made use of the NASA ADS database.
\end{acknowledgements}

\bibliographystyle{aa}
\bibliography{biblio}

\end{document}